\documentclass[twocolumn]{aastex63}
\usepackage{verbatim}
\usepackage{comment} 

\def\be{\begin{equation}}
\def\ee{\end{equation}}

\newcommand\quotes[1]{``{#1}"}
\def\gsim{\lower.5ex\hbox{\gtsima}} 
\def\lsim{\lower.5ex\hbox{\ltsima}} 
\def\gtsima{$\; \buildrel > \over \sim \;$} 
\def\ltsima{$\; \buildrel < \over \sim \;$} \def\gsim{\lower.5ex\hbox{\gtsima}} 
\def\lsim{\lower.5ex\hbox{\ltsima}} 
\def\simgt{\lower.5ex\hbox{\gtsima}} 
\def\simlt{\lower.5ex\hbox{\ltsima}}

\def\msun{{\rm M}_{\odot}}

\def\msunyr{\msun {\rm yr}^{-1}}

\def\S*{$\Sigma_{\rm SFR}$}

\def\kms{{\rm km\,s}^{-1}\,}

\definecolor{apcolor}{HTML}{b3003b}
\definecolor{afcolor}{HTML}{800080}
\definecolor{lvcolor}{HTML}{DF7401}
\definecolor{mdcolor}{HTML}{01abdf} 
\definecolor{cbcolor}{HTML}{ff0000}
\definecolor{sccolor}{HTML}{cc5500} 
\definecolor{sgcolor}{HTML}{00cc7a}

\makeatletter
\def\@hex@@Hex#1%
 {\if a#1A\else \if b#1B\else \if c#1C\else \if d#1D\else
  \if e#1E\else \if f#1F\else #1\fi\fi\fi\fi\fi\fi \@hex@Hex}
\makeatother
\definecolor{afcolor}{HTML}{b3443c}

\definecolor{apcolor}{HTML}{b3003b}

\shorttitle{Super-early massive galaxies}
\shortauthors{Ferrara et al.}
\graphicspath{{./}{figures/}}

\begin{document}

\title{On the stunning abundance of super-early, massive galaxies revealed by JWST}

\correspondingauthor{Andrea Ferrara}
\email{andrea.ferrara@sns.it}

\author[0000-0002-9400-7312]{Andrea Ferrara}
\affil{Scuola Normale Superiore,  Piazza dei Cavalieri 7, 50126 Pisa, Italy}

\author[0000-0002-7129-5761]{Andrea Pallottini}
\affil{Scuola Normale Superiore,  Piazza dei Cavalieri 7, 50126 Pisa, Italy}

\author[0000-0001-8460-1564]{Pratika Dayal}
\affil{Kapteyn Astronomical Institute, University of Groningen, 9700 AV Groningen, The Netherlands}

\begin{abstract}
The earliest JWST observations have revealed an unexpected abundance of super-early ($z>10$), massive ($M_*\, \approx 10^9 \msun$) galaxies at the bright-end ($M_{\rm UV}\approx -21$) of the ultraviolet luminosity function (UV LF). We present a \textit{minimal} physical model that explains the observed galaxy abundance at $z=10-14$. The model primarily combines (a) the halo mass function, with (b) an obscured star formation fraction prescription that is consistent with findings of the ALMA REBELS dusty galaxy survey. It has been successfully tested on well-known UV LFs up to $z=7$. The weak evolution from $z=7$ to $z\approx 14$ of the LF bright-end arises from a conspiracy between a decreasing dust attenuation, making galaxies brighter, that almost exactly compensates for the increasing shortage of their host halos. The model also predicts that galaxies at $z\simgt 11$ should contain negligible amounts of dust. We speculate that dust could have been efficiently ejected during the very first phases of galaxy build-up. \end{abstract}


\section{Introduction} \label{sec:intro}
Hierarchical models of galaxy formation based on the standard $\Lambda$CDM cosmological model have been very successful in predicting the properties and evolution of galaxies across a wide range of redshifts (for a recent review, see \citealt{Dayal18}). 

A number of theoretical approaches, ranging from numerical simulations \citep[e.g.][]{vogelsberger2014, schaye2015, roshdal:2018, pillepich:2018, hopkins:2018, dave:2019, Trebitsch2021, Pallottini22} to semi-analytic models \citep[e.g.][]{Dayal14, sommerville:2015, Lacey2016, Mutch2016, Behroozi19, Dayal22} have now been developed to study the formation and evolution of galaxies through cosmic time. 

These have been invaluable in shedding light on issues ranging from the (feedback-driven) assembly, composition (stellar, dust and black hole masses) and kinematics of early galaxies to their interplay with large-scale processes including reionization. However, observational data is crucial in both base-lining and testing the validity of such models.

Observationally, most of our statistical knowledge of galaxy populations in the Epoch of Reionization (redshift $z>6$) has, so far, been acquired by analyzing the UV Luminosity Function (LF) collected by the Hubble space Telescope (HST), \textit{Spitzer} and ground-based telescopes. These have been now reliably determined up to $z\simeq 9$ \citep{Bouwens21}, and have allowed the derivation of key quantities such as the evolution of the cosmic stellar mass density and star formation rate density (SFRD), among others. These data are constantly augmented and complemented by targeted Atacama Large Millimetre Array (ALMA) far-infrared observations. The recently completed ALMA REBELS program \citep{Bouwens21, Inami22} has firmly established that galaxies at $z\approx 7$ are already dust- and metal-enriched to unexpected levels \citep{Ferrara22}. This discovery has raised thorny questions on the amount of star formation that is missed due to dust obscuration \citep[][]{Fudamoto21}. Such effect dramatically impacts on LF and SFRD determinations, particularly at the highest masses, where we do expect dust extinction to be more severe \citep{Dayal22,Sommovigo22}.

The \textit{James Webb Space Telescope} (JWST), is already starting to revolutionize the field. However, as it is often the case, better observations pave the way for even more puzzling questions. Among the earliest results \citep{Naidu22, Donnan22, Finkelstein22, Atek22, Borsani22, Castellano22, Santini22, Adams22}, obtained via the public JWST Early Release Science programs (CEERS and GLASS), is the detection of two very bright ($M_{\rm UV} \simeq -21$) galaxies at super-early epochs \citep{Naidu22}: GLASS-z13 (GL-z13, photometric $z= 13.1^{+0.8}_{-0.7} $) and GLASS-z11 (GL-z11, $z = 10.9^{+0.5}_{-0.4}$); their stellar mass is $M_\star\simeq 10^9 \msun$. 

At face value this detection poses a serious challenge to essentially all models, and perhaps to the $\Lambda$CDM one itself. In fact, from the observed UV LF the estimated joint abundance of these two sources, $\log \phi\, [\rm Mpc^{-3}\rm mag^{-1}]= -5.05^{+0.47}_{-0.45}$, is $>10\times$ larger than predicted by models. This finding confirms (and strongly exacerbates) the tension between observations of the bright-end of the LF and theoretical predictions, which was already hinted at $z=7-9$ by \citet[][]{Bowler20, Harikane2022}. 

A large, unexpected abundance of bright galaxies at the earliest redshifts probed by JWST is a common finding among the above mentioned independent studies. In spite of the many uncertainties still present in the data analysis, it is therefore urgent to theoretically clarify such issues. 

In this \textit{Letter} our goal is to provide an interpretation to the unexpected abundance of super-early massive galaxies revealed by JWST, by using a simple LF evolution model.

Several possibilities can be envisaged and tested (see e.g. \citealt[][]{Mason22}), the most extreme of which is that the density of dark matter halos could be higher than predicted by $\Lambda$CDM. More likely, though, astrophysical factors play the key role. 

Among these are a very large conversion efficiency of the gas into stars at high-$z$, a top-heavy IMF or UV luminosity per stellar mass formed, up to the possibility of a magnification bias. Here we stick to the simplest (and perhaps most plausible) explanation: an (almost) vanishing dust obscuration beyond $z \simgt 11$. 
\section{Building the luminosity function}\label{sec:sims}
We aim at building a stripped-down, \textit{minimal} physical model of the redshift evolution of the UV LF, $\phi(M_{\rm UV})$. We start from the dark matter Halo Mass Function (HMF), ${\rm d}n(M,z)/{\rm d}\ln M$, i.e. the comoving number density of halos of mass $M$ at redshift $z$. We obtain the HMF by using the classical \citet{Sheth02} formalism, which uses an ellipsoidal model for perturbation collapse\footnote{Throughout the paper, we assume a flat Universe with the following cosmological parameters: $\Omega_{\rm M} = 0.3075$, $\Omega_{\Lambda} = 1- \Omega_{\rm M}$, and $\Omega_{\rm b} = 0.0486$,  $h=0.6774$, $\sigma_8=0.826$, where $\Omega_{M}$, $\Omega_{\Lambda}$, and $\Omega_{b}$ are the total matter, vacuum, and baryon densities, in units of the critical density; $h$ is the Hubble constant in units of $100\,\kms$, and $\sigma_8$ is the late-time fluctuation amplitude parameter \citep{planck:2015}.}. According to the HMF, at $z=11.5$ (the mean redshift of the two sources) galaxies with a number density of $\log n = -5.05\, \rm Mpc^{-3}$ are hosted by halos of mass $M=10^{11.33} M_\odot$. These objects are quite rare at those epochs, representing $4.8\sigma$ fluctuations of the density field. 

As a second step, we compute the galaxy (unattenuated) UV luminosity at 1500\AA, $L_{1500}$, as a function of $M$. This is proportional to the star formation rate (SFR) via a conversion factor\footnote{${\cal K}_{1500}$ has units of ${L_\odot}/(M_\odot {\rm yr}^{-1})$; its value has been chosen so to match the one used by the ALMA REBELS survey \citep{Bouwens22}, whose results will be used in the following.} ${\cal K}_{1500} \equiv {L_{1500}}/{\rm SFR} = 0.587 \times 10^{10}\,$. 

The simplest way to predict the SFR is given by the classical Schmidt-type expression,
\be\label{eq:SFR}
{\rm SFR} = \epsilon_* \left(\frac{\Omega_b}{\Omega_m}\right) \frac{M}{t_{\rm ff}},
\ee
where $\epsilon_*$ is the (in general, mass-dependent) star formation efficiency, and $t_{\rm ff} = (4\pi G \rho)^{-1/2}$ is the gas free-fall time in halos. 
The mean gas density within virialized structures is $\rho = 18\pi^2 \langle \rho \rangle$, where the mean cosmic density at redshift $z$ is $\langle \rho \rangle = \Omega_b (1+z)^3 \rho_{\rm cr} $, and the present-day critical density is $\rho_{\rm cr} = 3H_0^2/8\pi G$.

The free-fall time can be written as $t_{\rm ff}= \zeta H(z)^{-1}$, where $H(z)^{-1}$ is the Hubble time at $z$, and $\zeta=0.06$. We can then rewrite eq. \ref{eq:SFR} as
\be\label{eq:SFR1}
{\rm SFR} = \left(\frac{\epsilon_*}{\zeta}\right) \left(\frac{\Omega_b}{\Omega_m}\right) H_0  E(z) M
\ee
where $E(z)\approx \Omega_m^{1/2}(1+z)^{3/2}$ being a very good approximation at the high-$z$ of interest here; numerically,  
\be\label{eq:SFR1}
{\rm SFR} = 22.7 \left(\frac{\epsilon_*}{0.01}\right)\left(\frac{1+z}{8}\right)^{3/2} M_{12}\quad\msunyr\,,
\ee
with $M=10^{12}M_{12}\, \msun$. Eq. \ref{eq:SFR1} predicts that the SFR of GL-z11/GL-z13 should be 8.84 $\msunyr$, in excellent agreement with their observed values $(12^{+9}_{-4}, 7^{+4}_{-3})\, \msunyr$.  

Although not strictly necessary, as here we are concerned with the evolution of LF bright end, we include an additional ingredient, supernova (SN) feedback, to obtain the best possible agreement with the observed $M_{\rm UV}$ range. Feedback suppresses star formation in small galaxies by reducing $\epsilon_*$ in these systems. We adopt the physically motivated form, successfully tested on a wide redshift range, proposed by \citet{Dayal14}:
\be\label{eq:eps}
\epsilon_* = \epsilon_0\, \frac{v_c^2}{v_c^2+ f_w v_s^2}  
\ee
where $v_c(M)$ is the halo circular velocity, $f_w=0.1$ is the SN energy coupling efficiency with the gas, and $v_s = \sqrt{\nu E_0} = 975\, \kms$ is a characteristic velocity associated with the SN energy released per unit stellar mass formed: $E_0=10^{51}\, \rm erg$, $\nu^{-1} = 52.89\, \msun$, appropriate for a $1-100\,\msun$ Salpeter IMF. We fix $\epsilon_0=0.02$, consistent with measurements in local galaxies \citep[][]{Krumholz17}.

The luminosity associated with the SFR in eq. \ref{eq:SFR1} can be converted in a UV AB magnitude at 1500\AA:
\be\label{eq:MUV}
M_{\rm UV} = - 2.5\log L_{1500} + 5.89 - 1.087 \tau_{\rm eff}. 
\ee
The last term accounts for extinction, quantified by the \textit{effective} optical depth at 1500\AA, $\tau_{\rm eff}$\footnote{We emphasise the distinction between the optical depth physical value, $\tau_\lambda$, and the effective one $\tau_{\rm eff}$ (eq. \ref{eq:fobs}, \citealt[see, e.g.][]{Ferrara22}) which includes radiative transfer effects. For reference, the 1500\AA\, to V-band dust optical depth conversion is $\tau_{1500} = (2.655, 5.319) \tau_V$ for (Milky Way, SMC) extinction curves, respectively.}. Using the chain rule to transform $dn/dM$ into $dn/dM_{\rm UV}$, we can conveniently compare the predicted $\phi(M_{\rm UV})$ with data.

The only quantity that is still missing at this stage is the functional form of $\tau_{\rm eff}$, describing dust attenuation effects. In the next Section we describe this final step.  

\section{Dust attenuation}\label{sec:dust}

\begin{figure}
\centering\includegraphics[width = 0.45 \textwidth]{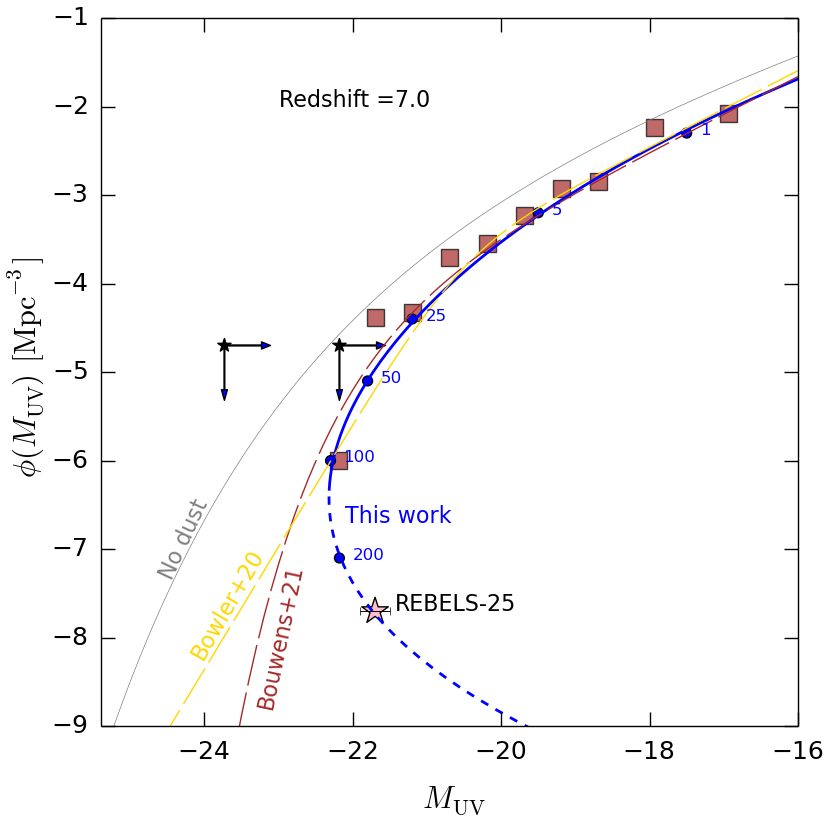}
\caption{Comparison between the predicted $z=7$ UV LF (blue curve) with data (red squares) and their best fit (red dashed curve) based on a Schechter function \citep[][]{Bouwens21}. The best fit using a double power-law \citep{Bowler20} is shown by the yellow dashed curve. Dots and numbers on the blue curve indicate the corresponding SFR (in $\msunyr$) of galaxies at a given $M_{\rm UV} - \phi$ location. The dashed portion of the blue curve, although unphysical in terms of the LF, is left to elucidate the abundance of obscured galaxies (see text). We also mark the $M_{\rm UV}=-21.7$ \citep[][]{Bouwens22} magnitude of the brightest, heavily obscured REBELS-25 galaxy \citep[][]{Inami22}, which is then predicted to have a number density $\log \phi =-7.7$. Also shown with black stars are two UV undetected galaxies at $z\approx 7$, REBELS-12-2 (left) and REBELS-29-2 (right) \citep{Fudamoto21}.
\label{Fig:01}
}
\end{figure}

We start by comparing our predictions to the $z=7$ UV LF. We choose this specific redshift as a benchmark because: (i) it provides a good compromise between the need to consider early cosmic epochs, and having a robust LF determination; (b) it allows us to exploit the ALMA REBELS to pivot and calibrate the dust attenuation. The results are shown in Fig. \ref{Fig:01}, where we also show the best fits to the observed LF at $z=7$ obtained by two independent studies: \citet[][]{Bouwens21}, who use a Schechter function, and \citet{Bowler20} adopting instead a double power-law. 

\begin{figure}
\centering\includegraphics[width = 0.45 \textwidth]{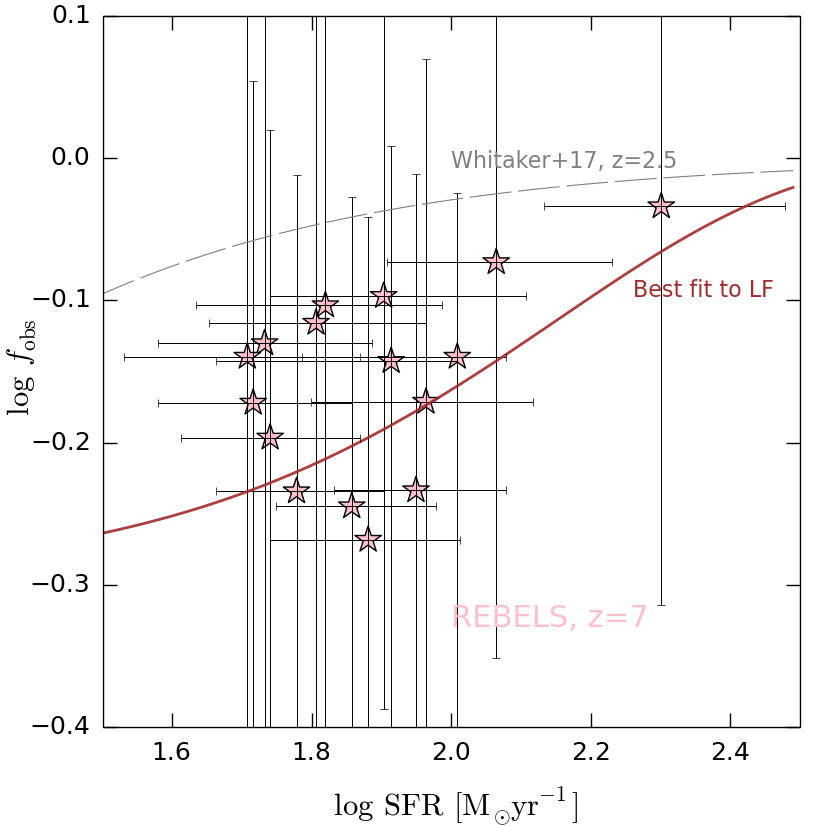}
\caption{Predicted dependence of the obscured fraction, $f_{\rm obs}= 1 - e^{-\tau_{\rm eff}}$ (red curve) yielding the best fit to the $z=7$ LF, compared to the $z=7$ REBELS data (pink stars, \citealt[][]{Bouwens22}). Also shown for reference is the fit for low ($z=2.5$) redshift galaxies by \citet{Whitaker17}.
\label{Fig:02}
} 
\end{figure}

As our model does not make predictions on dust attenuation we leverage on the recent results from the ALMA REBELS survey which showed that $\tau_{\rm eff}$ increases with the galaxy SFR \citep[][Fig. 4]{Ferrara22}. Hence we adopt a two-step procedure: (a) we find the functional form $\tau_{\rm eff}$(SFR) that best fits the LF at $z=7$; (b) we check that the derived relation is consistent with the obscured SFR fraction, 
\be\label{eq:fobs}
f_{\rm obs} = \frac{\rm SFR_{IR}}{\rm SFR_{UV}+SFR_{\rm IR}} = 1 - e^{-\tau_{\rm eff}},
\ee
inferred from the REBELS data \citep{Inami22, Ferrara22, Dayal22}. Such procedure yields the following relation:
\be\label{eq:taueff}
\tau_{\rm eff} = 0.7+0.0164\,\left(\frac{\rm SFR}{10\, \msunyr}\right)^{1.45}\,,
\ee
which is strictly valid for SFR $\simgt 1\, \msun$. In Fig. \ref{Fig:02} we compare eq. \ref{eq:fobs} with data points. Although errors are still large, the curve is consistent with the REBELS $f_{\rm obs}$, and properly catches the increasing trend of $f_{\rm obs}$ with SFR. We consider this as a successfully passed sanity check. 
%
%
%
%
%
%
\section{Model implications}\label{sec:implic}
Armed with our predicted LF at $z=7$, we compare it with the fits by \citet[][]{Bouwens22} and \citet[][]{Bowler20}. We also plot for reference the curve \quotes{No dust} in which dust attenuation effects are neglected ($\tau_{\rm eff}=0$). Our model is in perfect agreement with both fits for $M_{\rm UV} \simgt -22$, as seen in Fig. \ref{Fig:01}, but it differs for galaxies brighter than this magnitude.

We note that the model UV magnitude reaches a minimum at $M_{\rm UV}=-22.3$, corresponding to a SFR $\simeq 130\, \msunyr$ (SFR is marked by points along the blue model curve in Fig. \ref{Fig:01}). This is because galaxies with an intrinsically higher SFR are more attenuated (see eq. \ref{eq:taueff}), their magnitude decreases, and hence contribute to a fainter bin of $M_{\rm UV}$. Obviously, the number density of these highly obscured galaxies should be added to the upper branch of the bi-valued curve, but their number density is negligible so the actual LF (upper branch) is basically unaffected by their presence. Hence, the first model prediction is that galaxies brighter than $-22.3$ at $z=7$ should then be, at least partly, powered by an AGN, which seems consistent with GOLDRRUSH results \citep{Ono2018}, indicating an AGN contribution $\gsim 90\%$ at $M_{\rm UV}< -23$ already at $z=6-7$ \citep[see also ][]{Piana2022}.

The second prediction concerns the abundance of highly obscured galaxies. To illustrate this point, in Fig. \ref{Fig:01} we show the location of REBELS-25 ($M_{\rm UV}=-21.7$), the brightest and most heavily obscured REBELS galaxy \citep[][]{Inami22}. At that magnitude, REBELS-25 can be placed either on the upper (SFR $ =43\, \msunyr$) or lower ($267\, \msunyr$) part of the LF. The upper solution must be discarded as the total SFR, primarily driven by its large IR Luminosity, $\log (L_{\rm IR}/{L_\odot}) = {12.45}^{+0.43}_{-0.45}$ \citep{Sommovigo22}, has been estimated to be $200^{+101}_{-64} \msunyr$.
According to this picture, REBELS-25 is attenuated by about  $\approx 3$ magnitudes (the difference between the \quotes{No dust} curve, and the predicted LF at number density $\log \phi =-7.7$ in Fig. \ref{Fig:01}). 
We conclude that LIRGs like REBELS-25 are rare objects with a number density $\log \phi =-7.7$, which corresponds to 1/630 of all unobscured galaxies at the same UV mag. By evaluating the SFR contribution of obscured vs. unobscured systems, we conclude that galaxies like REBELS-25 only contribute $\simlt 1$\% of the cosmic SFRD. 

\section{Super-early massive galaxies}\label{sec:early}

We are now ready to interpret the recent LF measurements at $z = 11.5$. We show that the apparent lack of evolution in the bright-end of the UV LF is driven by a decreasing attenuation (i.e. a larger transmission of UV photons) compensating for a decreasing host halo abundance towards higher redshifts. We then make the hypothesis that the observed galaxies at $z=11.5$ are virtually obscuration-free.  

\begin{figure}
\centering\includegraphics[width = 0.45 \textwidth]{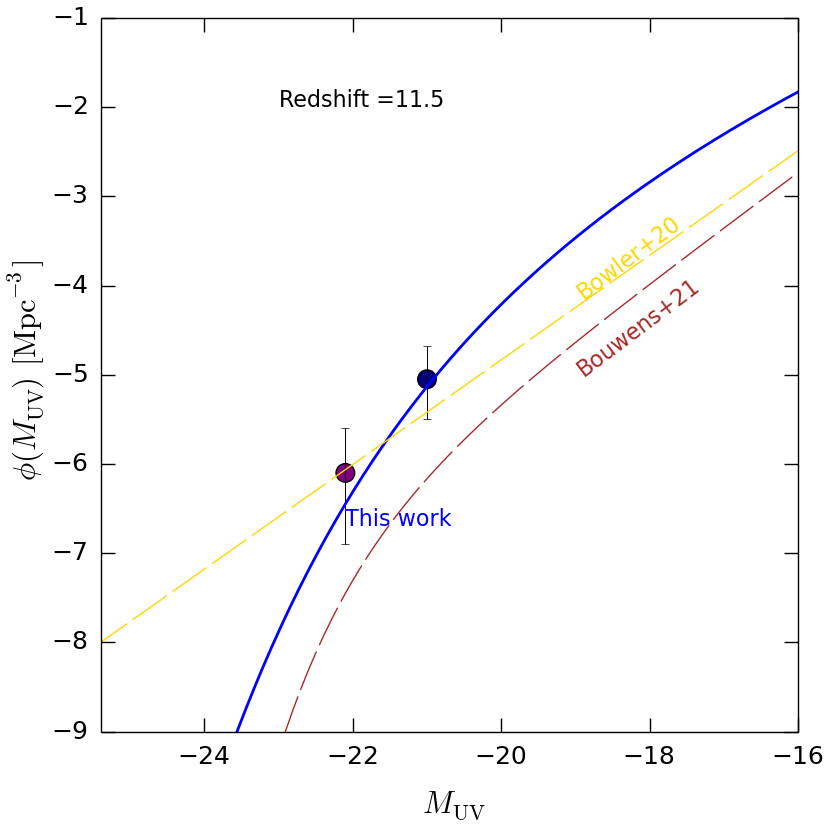}
\caption{Comparison between the predicted $z=11.5$ luminosity function and extrapolations of a lower redshift best fit based on a Schechter function \citep[][red dashed]{Bouwens21}, or a double power-law \citet[][yellow dashed]{Bowler20}. The blue point indicates the \citet{Naidu22} combined GL-z11/GL-z13 galaxies, while the brown point is the galaxy GN-z11 \citep[][]{Oesch16} at $z=11.1$.}
\label{Fig:03}
\end{figure}

The predicted LF, assuming $\tau_{\rm eff}=0$, is shown in Fig. \ref{Fig:03}, and it correctly matches the estimated number density of GLz11/GLz13. The solution also successfully reproduces the data point for GN-z11 \citep{Oesch16}, a spectroscopically confirmed galaxy at $z=11$. Interestingly, though, our solution is about a factor 10 above the extrapolated \citet{Bouwens21} Schechter fit; it is instead consistent with the \citet{Bowler20} double-power law extrapolation. However, we note that the shape of our LF is neither a Schechter nor a double power-law function.

As a further test of our predictions, we compute the LF at $z=13.75$, so to be able to compare it with another set of JWST detections by \citet{Donnan22}. In their Fig. 2, they show a single point of the LF in the highest redshift bin $12.5 < z < 15$ (mean $z=13.75$) containing 5 sources identified from the JWST ERO and ERS NIRCam imaging (SMACS0723, GLASS, CEERS) in combination with deep ground-based near-infrared imaging in the COSMOS field. The measurement at $M_{\rm UV}=-19.1$ is $\log \phi = -4.77^{+0.24}_{-0.53}$. Our model, assuming $\tau_{\rm eff}=0$ as at $z=11.5$, consistently predicts $\log \phi = -4.1$, i.e. within $\approx 1\sigma$ from the observed value. 

These findings lead us to conclude that recently discovered super-early massive galaxies are virtually unaffected by dust attenuation, i.e. $\tau_{\rm eff} \simeq 0$. This might be due to a special relative distribution of stars and dust, allowing substantial 1500\,\AA\, radiation leakage. Alternatively, these systems might have a particularly low dust-to-stellar mass ratio. A possibility to be explored is that dust during these very early evolutionary stages is promptly evacuated by radiation pressure as soon as it is produced by stars (Ziparo et al., in prep.). In both cases, though, we expect these systems to be poor FIR continuum emitters. 

Tantalisingly, extremely low attenuation is reported for the highest redshift candidates discovered by JWST. For example, Maisie's galaxy ($z\approx 14, M_{\rm UV}=-20.3, {\rm SFR}=4.1\, \msunyr, M_*=10^{8.5}\ \msun$ has $A_V=0.06$ \citealt{Finkelstein22}). The highest-$z$ candidate (GHZ2, $z=12.35, M_{\rm UV}=-21.9$) of the two found by the GLASS program (the other being GHZ1, $z=10.6$) shows a very blue slope ($\beta = -2.43 \pm 0.11$, \citealt{Castellano22, Santini22}). Similarly, the two highest redshift candidates found by \citet{Atek22}, SMACS-z12a ($z=12.03$) and SMACS-z12b ($z=12.35$) have extremely steep UV spectral slopes, $\beta=-2.60,-2.71$, respectively. Finally, the galaxy 10234 at $z=11.49$ found by \citet{Adams22} in the ERO SMACS 0723 field, is the bluest (with a striking $\beta \approx -3.35$) among the four $z>9$ candidates. All these galaxies are very massive for their epoch (stellar masses $M_*=10^{8-9} \msun$). Hence, one would have naively expected a significant dust attenuation. Although the above examples do not constitute a proof of an overall highly decreased attenuation at $z \simgt 11$, such a coincidence is nevertheless hard to be overlooked. 

Beside dust, alternative explanations might -- in principle -- account for the observed abundance of super-early, bright galaxies. The first is the magnification bias due to (weak and strong) gravitational lensing \citep[][]{Wyithe11, Mason15}. However, these studies have shown that this effect is significant only for $M_{\rm UV} < -22$. A second possibility is an IMF that becomes increasingly top-heavy with increasing redshift. This has been advocated to simultaneously match both the number counts and redshift distribution of sub-mm galaxies \citep[][]{Lacey16}. Whether these are viable solutions is left to further work.
%
%
%

%
%
\begin{figure}
\centering\includegraphics[width = 0.45 \textwidth]{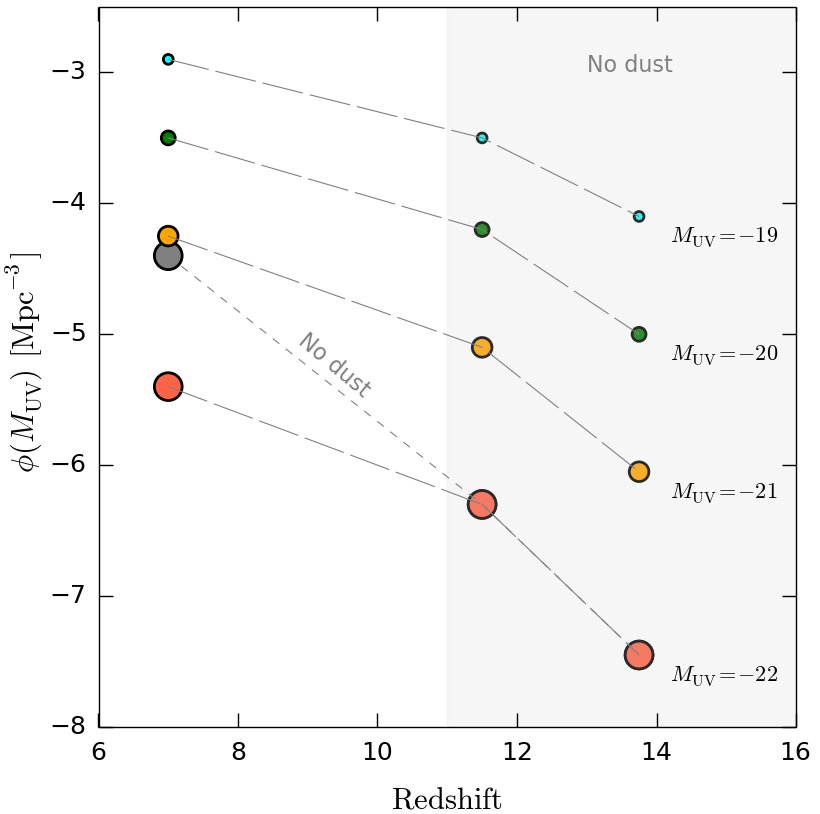}
\caption{Predicted luminosity density evolution for three relevant redshifts ($z=7, 11.5, 13.75$) discussed in this work. Each curve refers to a different $M_{\rm UV}$ as shown by the labels. The grey point on the $M_{\rm UV}=-22$ line assumes $\tau_{\rm eff}=0$; in this case the evolution of $\phi(M_{\rm UV})$ would be much steeper than observed. The model assumes that beyond $z\approx 11$ (gray vertical band) dust attenuation is negligible. Hence, all the points shown in that area do not include dust attenuation.}
\label{Fig:04}
\end{figure}

\section{Luminosity function evolution}\label{sec:lfevo}

From the above results we build the redshift evolution of the LF (Fig. \ref{Fig:04}). The number density $\phi(M_{\rm UV})$ at fixed $M_{\rm UV}$ decreases with redshift. However, the decrease becomes much steeper at $z \simgt 11$, where we have assumed no dust attenuation. The effect of a decreasing dust attenuation/content, as already stated, is to compensate the rapid density drop, otherwise imposed by the underlying HMF, via a galaxy luminosity enhancement. 

To illustrate this point, in Fig. \ref{Fig:04} we also show the (artificial) case $\tau_{\rm eff} = 0$ for $z=7$. If dust effects can be neglected beyond $z \approx 11$, then we do expect an accelerated drop of $\phi(M_{\rm UV})$ that is particularly evident at bright magnitudes. Encouragingly, such accelerated evolution is consistent with the number density by \citet[][]{Donnan22} at $z=13.75$ (see Sec. \ref{sec:implic}).    

\section{Summary}\label{sec:summary}

A minimal galaxy evolution model including HMF, supernova feedback, and dust attenuation calibrated on the ALMA REBELS results correctly predicts the LF at $z=7$. The main predictions are:

\begin{itemize}
\item[{\color{red} $\blacksquare$}] The effective optical depth $\tau_{\rm eff} \propto {\rm SFR}^{1.45}$. This implies that galaxies brighter than $M_{\rm UV} = -22.3$ at $z = 7$ should have a significant contribution from AGN. 

\item[{\color{red} $\blacksquare$}] 
LIRG-like galaxies as REBELS-25,  $\log (L_{\rm IR}/{L_\odot}) = {12.45}$, contribute $\simlt 1$\% of the cosmic SFR density.

\item[{\color{red} $\blacksquare$}] 
The model correctly predicts the abundance of super-early, bright JWST candidates both at $z=11.5$ \citep[][]{Naidu22} and $13.75$ \citep[][]{Donnan22}, by \textit{assuming a negligible dust attenuation at those redshifts}.

\item[{\color{red} $\blacksquare$}] 
The no-attenuation hypothesis is supported by the very blue UV slopes $\beta \simlt -2.5$ of the highest redshift candidates identified by four different JWST studies. 

\item[{\color{red} $\blacksquare$}] 
We speculate that dust could have been efficiently ejected during the earliest phases of galaxy build-up now accessible thanks to JWST.

\item[{\color{red} $\blacksquare$}] 
The weak evolution from $z=7$ to $z\approx 14$ of the LF bright end arises from the conspiracy between a decreasing dust attenuation, making galaxies brighter, that almost exactly compensates for the increasing shortage of their host halos.

\end{itemize}

\section*{Data Availability}
Data available on request.

\acknowledgments
We thank S. Carniani, P. Rosati, E. Vanzella, F. Ziparo for insightful comments, and M. Castellano for providing data on GHZ2.
AF, AP acknowledge support from the ERC Advanced Grant INTERSTELLAR H2020/740120. Generous support from the Carl Friedrich von Siemens-Forschungspreis der Alexander von Humboldt-Stiftung Research Award is kindly acknowledged (AF). 
PD acknowledges support from the NWO grant 016.VIDI.189.162 (``ODIN") and from the European Commission's and University of Groningen's CO-FUND Rosalind Franklin program.
Plots in this paper produced with the \textsc{matplotlib} \citep{Hunter07} package for \textsc{PYTHON}.    

\bibliographystyle{aasjournal}
\bibliography{paper}



\end{document}